\documentclass[twocolumn,prl, showpacs]{revtex4-1}
\usepackage{graphicx}% Include figure files
\usepackage{dcolumn}% Align table columns on decimal point
\usepackage{bm}% bold math
\usepackage{color}%colorful font
\usepackage{appendix}

\begin{document}

%\preprint{}

\title{Graphene-like conjugated $\pi$ bond system in Pb$_{1-x}$Sn$_x$Se}
% in thermodynamic canonical ensemble
% PSS-(001) monoatomic layer as a square-graphene ?
%\\ as a thermoelectric material to topological crystalline insulator 

\author{G. J. Shu$^1$}
\author{S. C. Liou$^2$}
\author{S. Karna$^1$}
\author{R. Sankar$^1$}
\author{M. Hayashi$^1$}
\author{M.-W. Chu$^1$}
\author{F. C. Chou$^{1,3,4}$}
\email{fcchou@ntu.edu.tw}
\affiliation{
$^1$Center for Condensed Matter Sciences, National Taiwan University, Taipei 10617, Taiwan}
\affiliation{
$^2$NISP Lab, Nano Center, University of Maryland, College Park, MD 20742, U.S.A. }
%\affiliation{
%$^3$Japan}
%\affiliation{
%$^3$Department of Physics, Boston College, Chestnut Hill, Massachusetts 02467}
\affiliation{
$^3$National Synchrotron Radiation Research Center, Hsinchu 30076, Taiwan}
\affiliation{
$^4$Taiwan Consortium of Emergent Crystalline Materials, Ministry of Science and Technology, Taipei 10622, Taiwan}
%\affiliation{
%$^3$Center for Emerging Material and Advanced Devices, National Taiwan University, Taipei 10617, Taiwan}

\date{\today}

\begin{abstract}
Following the identification of the $\pi$ bond in graphene, in this work, a $\pi$ bond constructed through side-to-side overlap of half-filled $6p_z$ orbitals was observed in a non-carbon crystal of Pb$_{1-x}$Sn$_x$Se (x$\sim$0.34) (PSS), a prototype topological crystalline insulator (TCI) and thermoelectric material with a high figure-of-merit ($ZT$).  PSS compounds with a rock-salt type cubic crystal structure was found to consist of $\sigma$ bond connected covalent chains of Pb(Sn)-Se with an additional $\pi$ bond that is shared as a conjugated system among the four nearest neighbor Pb pairs in square symmetry within all $\{$001$\}$ monoatomic layers per cubic unit cell.  The $\pi$ bond formed with half-filled $6p_z$ orbitals between Pb atoms is consistent with the calculated results from quantum chemistry.  The presence of $\pi$ bonds was identified and verified with electron energy-loss spectroscopy (EELS) through plasmonic excitations and electron density (ED) mapping via an inverse Fourier transform of X-ray diffraction. 
\end{abstract}


%79.20.Uv    Electron energy loss spectroscopy 
%71.15.Ap  Valence-bond method in electronic structure of solids
% 87.59.-e    X-ray imaging, 
%61.50.Ah	Theory of crystal structure, crystal symmetry; calculations and modeling

%73.20.-r	  Electron states at surfaces and interfaces
%75.10.Pq   Spin chain models, 
%61.05.cp	X-ray diffraction
%73.20.At	       Surface states, band structure, electron density of states
%73.25.+i	    Surface conductivity and carrier phenomena
%31.10.+z   Theory of electronic structure, electronic transitions, and chemical binding
%33.15.Fm	Bond strengths, dissociation energies
%72.20.-i	     Conductivity phenomena in semiconductors and insulators
%72.25.Dc	Spin polarized transport in semiconductors
%72.80.Vp	Electronic transport in graphene
%73.22.Pr	    Electronic structure of graphene
%75.70.Tj	    Spin-orbit effects (see also 71.70.Ej Spin-orbit coupling, Zeeman and Stark splitting, Jahn-Teller effect)

\maketitle

%\section{\label{sec:level1} Introduction\protect\\ }

%\textcolor{red}{Note also that the cubic symmetry protection is the consequence of the requirement of max potential energy reduction in Gibbs free energy form.}

In search of high $ZT$ compounds for thermoelectric applications, an intricate balance among the electric conductivity, thermal conductivity, and Seebeck coefficient is expected, which often leads to materials falling in a critical range between narrow band gap semiconductors and semimetals.\cite{Dresselhaus2007}  The latest findings on topological insulator (TI) and topological crystalline insulator (TCI) materials show that these also fall into nearly the same critical range.\cite{Muchler2013}  Extensive studies of TI and TCI materials have mostly investigated symmetry, and analysis has been performed from the perspective of energy-momentum in reciprocal space.\cite{Hasan2010}  A more intuitive interpretation from a real-space chemical bond model that shows the actual electron density distribution is desirable and would be especially useful for materials scientists and chemists focusing on materials design for future high $ZT$ thermoelectric and TCI materials.  

\begin{figure}
\begin{center}
\includegraphics[width=3.5in]{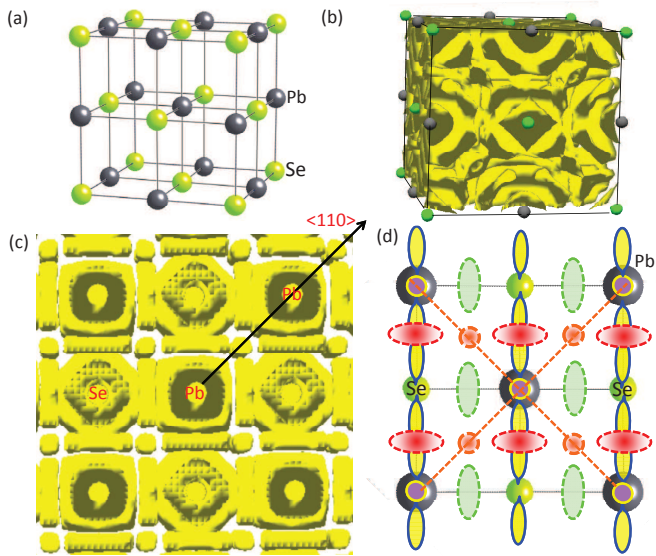}
\end{center}
\caption{\label{fig-PbSeED} (color online) (a) Crystal structure and (b) electron density contour mapping of Pb$_{1-x}$Sn$_x$Se (x$\sim$0.34) in 3D.  (c) A thick cut of ED for the PSS-$\{$001$\}$ layer between z=0.33 and 0.83.  The circle-shaped electron clouds (sashed circle in orange) in the diagonal direction between Pb(Sn) atoms are proposed to arise from the $\pi$ bond electrons.  (d) The corresponding molecular orbital model for the PSS-$\{$001$\}$ monoatomic layer in $\langle$001$\rangle$ projection; details are shown in Fig~\ref{fig-PbSeMO}(c)-(d).  The four bar-shape electron clouds shown in (c) correspond to the electrons in $\sigma$ bond between Pb(Sn)-Se (dashed oval in red) and the lone pair electrons of Se (dashed oval in green) shown in (d); however, these two are not distinguishable for PSS in cubic symmetry.}
\end{figure}

PbTe-based materials exhibit a demonstrated high thermoelectric figure-of-merit ($ZT$) at high temperatures and are therefore good candidates for thermoelectric device applications.\cite{Dresselhaus2007, Korkosz2014}  Recently, Pb$_{1-x}$Sn$_x$Se and Pb$_{1-x}$Sn$_x$Te with cubic symmetry have also been predicted and confirmed to be topological crystalline insulators (TCIs) for $x$ within a critical narrow band gap range, where band inversion and spin-orbital coupling occur for the crystalline symmetry-protected topological orders.\cite{Fu2011, Xu2012, Dziawa2012, Okada2013, Neupane2014} However, for PSS compounds with a rock-salt type face-centered cubic (FCC) Bravais point lattice (Fig.~\ref{fig-PbSeED}(a)), several chemical and physical features have yet to be characterized.  First, PSS compounds of rock-salt type in FCC symmetry do not easily cleave along the close-packed $\{$111$\}$ planes with the largest inter-planar distance and, instead, always cleave along the $\{$001$\}$ planes.\cite{Xu2012, Okada2013, Wang2013}  This observation suggests that strong anisotropic chemical bonding must exist despite both Pb/Sn and Te/Se atoms exhibiting undistorted octahedral coordination within a cubic rock-salt structure.  Most surprisingly, surface mirror symmetry breaking along the $\langle$110$\rangle$ directions has been observed in recent high-resolution STM studies of a PSS compound in the critical TCI regime at low temperature,\cite{Okada2013} which creates a puzzling conflict with the persistent 3D cubic symmetry of the bulk.\cite{Neupane2014, Littlewood1980}  

Graphene has been shown to possess a conjugated $\pi$ bond system and a Dirac-cone-shaped surface band.\cite{Neto2009, Fuhrer2010}  In this study, we find that the $\pi$ bond can also be identified in the confirmed TCI material Pb$_{1-x}$Sn$_x$Se (x$\sim$0.34) with the presence of Dirac cones.\cite{Xu2012, Dziawa2012}  Direct correlation between the $\pi$ band and Dirac cone via pseudo-spin has been previously established for graphene,\cite{Fuhrer2010} which is consistent with the conjugated nature of the $\pi$ bond at the molecular orbital level.\cite{Chua2013}  To establish the intimate relationship between the conjugated $\pi$ bond system and the presence of the Dirac cone, especially when the Dirac cone implies a gapless semiconducting behavior, we must first demonstrate the existence of the conjugated $\pi$ bond in PSS.  The electronic structure of the PSS compound was explored using X-ray electron density mapping and electron energy-loss spectroscopy (EELS) in this study.  A molecular orbital chemical bond model is proposed to explain the observed experimental results, where PSS compounds are proposed to consist of covalent chains via $\sigma$-bond-connected Pb(Sn)-Se through hybridized $s$ and $p$ ($sp$) orbitals in the valence shell; moreover, conjugated $\pi$ bond condensation similar to that observed in graphene is proposed to exist between Pb along the $\langle$110$\rangle$ diagonal directions.

%\section{\label{sec:level1} Electron density mapping of $PSS$\protect\\}

%===========================================================
%============================================================

Electron density (ED) mapping has been used to determine the actual valence electron distribution between bonded atoms in condensed matter.  The real space electron contour  can be determined experimentally with the aid of an inverse Fourier transform of the reciprocal $k$-space information obtained via X-ray diffraction.  The theoretical foundation of applying the Fourier method in crystallography has been thoroughly documented.\cite{EDtheory}  The bonding nature and distribution of electrons within chemical bonds can be visualized using this technique.\cite{Takata1994, Karna2013}  Here, the ED of Pb$_{1-x}$Sn$_x$Se (x$\sim$0.34) has been extracted from the inverse Fourier transform of powder X-ray diffraction data using a step size of 0.1 \AA$^{-1}$ at room temperature, as illustrated in Fig.~\ref{fig-PbSeED}(b).   

A thick slice (z=0.33-0.83) of the ED for PSS is shown in Fig.~\ref{fig-PbSeED}(c).  In addition to the expected ED from electrons in $\sigma$ bonds between Pb(Sn) and Se along the $\langle$100$\rangle$-directions in cubic symmetry, it is puzzling to observe non-zero ED with a circular shape in the diagonal $\langle$110$\rangle$ directions between Pb(Sn) atoms.  We believe that these experimentally observed electron clouds are contributed by unexpected bonding electrons that link the four nearest neighbor Pb(Sn) atoms in square symmetry per $\{$001$\}$ layer.  To explore the character of such an unconventional chemical bond between Pb(Sn), EELS studies were performed.  

%\section{\label{sec:level1} EELS studies of $PSS$ \protect\\}

%===========================================================
\begin{figure}
\includegraphics[width=3.5in]{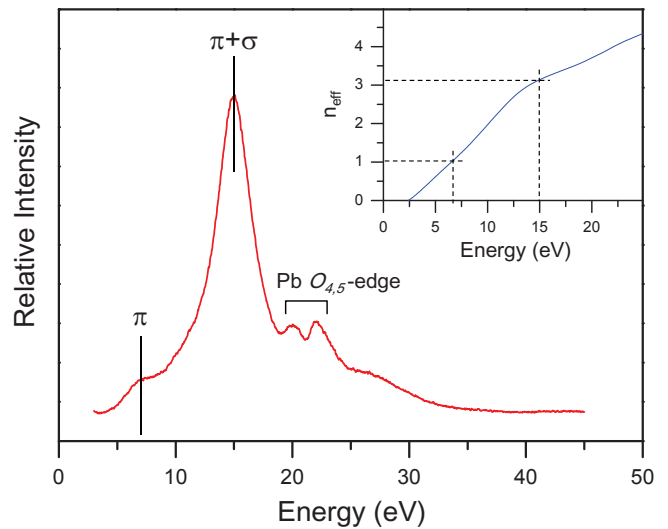}
\caption{\label{fig-EELS}(color online) EELS spectrum of Pb$_{1-x}$Sn$_x$Se (x$\sim$0.34).  The inset shows the $n_{eff}$ integrated from the energy-loss function, which indicates that the two peaks near 7 and 15 eV correspond to $n_\pi$:$(n_\pi$+$n_\sigma$)=1:(1+2) electrons per atom participating in the e-beam energy absorption from $\pi$ and $\pi$+$\sigma$ plasmons. The 20-22 eV weak peaks are due to the \textit{O$_{4,5}$}-edge excitation of Pb. }     
\vspace{-5mm}
\end{figure}
%============================================================

\begin{figure}
\begin{center}
\includegraphics[width=3.5in]{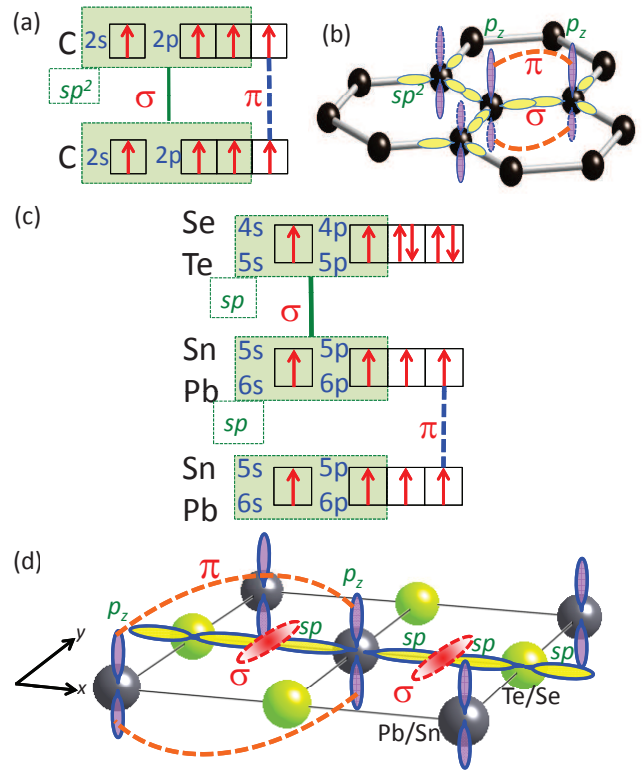}
\end{center}
\caption{\label{fig-PbSeMO}(color online) (a) The electronic configuration and (b) molecular orbital model of graphene, where the $\sigma$ bonds are formed with half-filled $sp^2$ hybrid orbitals per carbon in the honeycomb structure, and each $\pi$ bond is shared by the three nearest neighbor C-C pairs in both space and time. (c) The electronic configurations of Pb/Sn and Te/Se and (d) the corresponding molecular orbital model for a representative $\{$001$\}$ monoatomic layer of the PSS compound.  $\sigma$ bonds are formed with the half-filled hybridized $sp$ orbitals per Pb/Sn and Te/Se along any equivalent axis in cubic symmetry (assumed $x$-axis here). $\pi$ bonds are proposed to form within both the equivalent  $xy$- and $xz$-planes in cubic symmetry, as shown between the half-filled $p_z$  orbitals of Pb(Sn) atoms when $xy$-plane is chosen, through side-to-side orbital overlap along the $\langle$110$\rangle$ diagonal directions. }
\end{figure}

EELS can provide useful information regarding the effective number of electrons participating in the collective oscillations within chemical bonds through energy absorption of the $\pi$ or $\sigma$ plasmons.  EELS spectra for graphite and Bi$_2$Se$_3$ have been compared to provide evidence of the volume plasmons corresponding to the collective excitations of electrons within $\pi$ and $\sigma$ bonds.\cite{Liou2013}  The assignment of $\pi$ and $\sigma$ plasmons has been satisfactorily verified in graphite on the experimental values of the effective number of electrons per atom participating in the collective plasma oscillation, $n_{eff}(\omega)$, as $n_\pi$:$(n_\pi$+$n_\sigma$)=1:(1+3) for the two peaks in low-loss region of the EELS spectrum near 7 and 27 eV, respectively.\cite{Liou2013, Carbone2009} This finding can be interpreted as there being 1 and 3 electrons per carbon participating in the $\pi$ and $\sigma$ bonding, respectively;  these numbers are consistent with the molecular orbital model of three $\sigma$ bonds and one $\pi$ bond that is shared by three equivalent nearest neighbor carbon pairs in a honeycomb lattice construction, as illustrated in Fig.~\ref{fig-PbSeMO}(b).  The randomly distributed $\pi$ bond in space and time can be viewed as a conjugated system, similar to those observed in conductive polymers.\cite{Shirakawa2003} 

The EELS spectrum of Pb$_{1-x}$Sn$_x$Se (x$\sim$0.34) is shown in Fig.~\ref{fig-EELS}.  In parallel to the successful interpretation of EELS spectra for graphite and Bi$_2$Se$_3$,\cite{Liou2013, Carbone2009} the low-loss peaks near 7 and 15 eV can be assigned as corresponding to the energy absorption by the $\pi$ and $\pi$+$\sigma$ plasmons, respectively.  $\pi$-plasmon has also been revealed in the EELS spectrum of nanocrystal PbSe,\cite{Gunawan2014} and the two weak peaks at $\sim$20-22 eV are identified as the Pb \textit{O}$_{4,5}$-edge excitation from Pb $5d$ electrons within an octahedral coordination.  The effective number of electrons participating in the collective plasma oscillations of $\pi$ and $\pi$+$\sigma$ plasmons is determined by the energy-loss function analysis to be $\sim$1:3 corresponding to the two peaks near 7 and 15 eV, respectively, as shown in the inset of Fig.~\ref{fig-EELS}.  Following the similar assignment of $n_\pi$ and $n_\sigma$ for graphite,\cite{Liou2013} the effective number of electrons participating in the two EELS low-loss peaks for PSS near 7 and 15 eV, $\sim$1:3 (inset of Fig.~\ref{fig-EELS}), can be interpreted as $n_\pi$:$(n_\pi$+$n_\sigma$)=1:(1+2), i.e., there being 1 and 2 electrons per atom participating in the $\pi$ and $\sigma$ bonding, respectively.  Clearly, the proposed existence of $\pi$ bond in a non-carbon system PSS seems surprising, which requires a proper molecular orbital model to support.    

%\textcolor{blue}{Quote from Gunawan2014: resonant oscillations of the delocalized ¹ electrons, known as Ò$\pi$-plasmonsÓ}

%\section{\label{sec:level1} Molecular orbital model for $PSS$\protect\\ }

%===========================================================
%============================================================  

The crystal structure of PSS compounds has a rock-salt type cubic symmetry as shown in Fig.~\ref{fig-PbSeED}(a).  All of the atoms in PSS compounds can be described using an FCC Bravais point lattice with the selection of the (Pn/Sn)-(Te/Se) unit as a di-atomic basis, which is crystallographically identical to the typical NaCl ionic crystal.  Because both Pb/Sn and Te/Se atoms have octahedral coordination, conventionally the crystal structure is drawn assuming that chemical bonds exist for all six (Pb/Sn)-(Te/Se) pairs within each PbSe$_6$ or SePb$_6$ octahedron.  However, the electronegativity differences between Pb(2.33)/Sn(1.96) and Te(2.10)/Se(2.55) are rather small to suggest the existence of a polar covalent bond between Pb/Sn and Te/Sn based on Pauling's rule of ionicity.\cite{EN} 
%which is in great contrast to that of the typical ionic crystal of Na(0.93) and Cl(3.16) with a large EN difference of 2.23.  
The assumption of six chemical bonds per Pb/Sn within an octahedral coordination faces two contradictory facts; the first one is that the four valence shell electrons per Pb/Sn (Pb=[Xe]$4f^{14}5d^{10}6s^26p^2$) are insufficient to form six $\sigma$ bonds with the six neighboring Te/Se (Se=[Kr]$4d^{10}5s^25p^4$); and the second one is that for Te/Se in the chalcogen family below oxygen, at most two electrons are needed for stabilization into a filled inert gas configuration; i.e., a coordination of two is expected.  It is apparent that the conventional crystal structure drawn with chemical bonds to link all (Pb/Sn)-(Te/Se) in octahedral coordination could be heuristic and requires further examination.  

As a prototype IV series semiconductor, the chemical bond of Si crystal has been satisfactorily described through hybridized $sp^3$ orbitals for all Si atoms in tetrahedral coordination.  However, the same bonding theory failed to apply directly to the IV-VI compounds of rock-salt type cubic structure, such as PbSe-PbTe-SnTe, mostly because of the partially covalent polarity due to the small electronegativity difference and the unpaired $p$ orbital electrons.\cite{Nakanishi1980}  In parallel to the well-known molecular orbital model for graphene as shown in Fig.~\ref{fig-PbSeMO}(a)-(b), we propose a molecular orbital model that could maximize the number of chemical bond for PSS to reach the condensed matter ground state following the valence shell electron pair repulsion (VSEPR) theory rigorously.\cite{inorganic}  Under the requirement of a coordination number of two for Te/Se in the chalcogen family, it is natural to assume $sp$ hybridization from the $s$ and $p$ orbitals in valence shell near the Fermi level for both Pb/Sn and Te/Se, which leads to the formation of a covalent bond through the head-to-head overlap of half-filled $sp$ orbitals as a $\sigma$ bond, as illustrated in Fig.~\ref{fig-PbSeMO}(c)-(d).  These covalent $\sigma$ bonds form chains along one of the three equivalent crystal axes in cubic symmetry following nucleation.  However, the remaining two $p_i$ ($i=x,y,z$) orbitals for Te/Se are both filled as inactive lone paired electrons, whereas  the two remaining $p_i$ orbitals for Pb/Sn are both half-filled and subject to additional chemical bonding.  One of the most probable chemical bonds in addition to $\sigma$ bond formation could be a $\pi$ bond formed between the half-filled $6p_z$ orbitals of Pb atoms along the diagonal direction of $xy$-plane, as illustrated in Fig.~\ref{fig-PbSeMO}(d), where $\sigma$ bond chain is named along the $x$-axis within the $xy$-plane and the other equivalent choice of $\pi$ bond formation via $6p_y$ pairing within the $xz$-plane is not shown.  %The question next is whether it is possible for these half-filled $6p_{x'/y'}$ pairs of Pb atoms to form $\pi$ bonds in PSS compounds, similar to that of $\pi$ bonds formed with $2p_z$ orbitals of carbon atoms in graphene.

%===========================================================
\begin{figure}
\begin{center}
\includegraphics[width=3.5in]{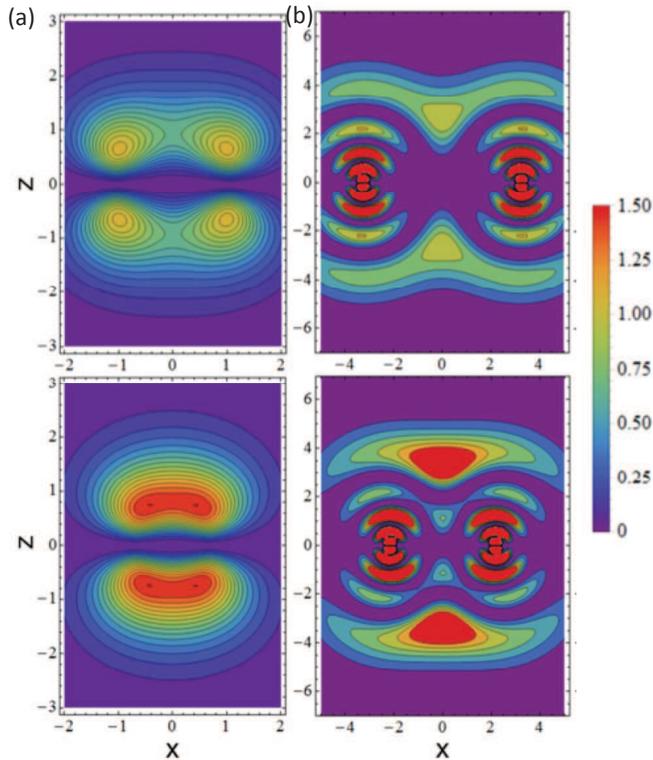}
\end{center}
\caption{\label{fig-2p6ppi}(color online) The calculated contours of the electron density distribution for two atoms at equilibrium interatomic distance (lower) and 1.5 times farther apart (upper).  (a) Two half-filled $2p_z$ orbitals between C-C atoms, (b) two half-filled $6p_z$ orbitals between Pb-Pb atoms.  The equilibrium interatomic distances between C-C of 1.4 $\AA$ is acquired from graphene, and the equilibrium distance between Pb-Pb is acquired from the PSS crystal structure of $\frac{\sqrt{2}}{2}a$$\sim$4.3$\AA$.  For presentation purpose, the charge densities of (a) C-C and (b) Pb-Pb are multiplied by factors of 5 and 100, respectively.  }
\end{figure}
%============================================================

Based on the $\pi$ bond formation via side-to-side overlap of half-filled $2p_z$ orbitals in graphene, we decide to explore the possibility of $\pi$ bond formation via side-to-side orbital overlap of half-filled $6p_z$ (see Fig.~\ref{fig-PbSeMO}(c)-(d)) orbitals between Pb.  Starting from the calculation of the probability function ($|\Psi(x,y,z)|^2$) of the corresponding $n\pi$ bond orbital wave functions $\Psi_{n\pi}(x,y,z)$=$N(\psi_{n10}+\psi_{n^\prime 10}$), where $\psi_{n10}$ is an $np_z$ orbital function of the atom of interest, a comparative study of the electron density distribution between $2p_z$ and $6p_z$ pairs as a function of inter-atomic distance is presented for C-C and Pb-Pb pairs in Fig.~\ref{fig-2p6ppi}.  The two atoms are located along the $x$-axis and the Slater rule is adopted for each atom to construct the $np_z$ wave function.  Fig.~\ref{fig-2p6ppi}(a) shows the calculated contours in the $y=0$  plane section of the electron density of two carbon atoms with C-C at variable distances.  It is apparent that a $\pi$ bond is formed by the two half-filled $2p_z$ orbitals when the inter-carbon distance is close to the average C-C bond length ($\sim$1.4 \AA) for graphene.\cite{Neto2009}  Interestingly, for the PSS crystal with equilibrium Pb-Pb distance of $\frac{\sqrt{2}}{2}a$$\sim$4.3 \AA, Fig.~\ref{fig-2p6ppi}(b) shows that a $\pi$-bond-like electron distribution can also form between the two half-filled $6p_z$ orbitals of Pb atoms with an inter-atomic distance of $\sim$4.3 \AA.  While the $2\pi$ bond orbital has no node along the $z$-direction and $6\pi$ has multiple nodes with nonzero electron density near each nucleus, the outer shell of $6\pi$ does form continuous electron density between the $6p_z$ pair resembling the typical $2\pi$ bond orbital.  

Considering the two unpaired electrons in each hybridized $sp$ orbital of Pb(Sn) and Se, and the one unpaired electron within $p_i$ ($i=y,z$) orbitals of Pb(Sn) (see Fig.~\ref{fig-PbSeMO}(c)-(d) when $\sigma$ bond is named along the $x$-axis), each $\{$001$\}$ monoatomic layer has two unpaired electrons in $sp$ orbital per Pb(Sn) and Se participating in the $\sigma$ bonding, and one unpaired electron in $p_{z}$ or $p_{y}$ (depending on the choice of $xy$- or $xz$-plane, respectively) per Pb(Sn) participating in the $\pi$ bonding, which is in perfect agreement with the interpretation of EELS spectrum of PSS (Fig.~\ref{fig-EELS}) showing that $n_\sigma$ $\approx$ 2 for $\sigma$ plasmon and $n_\pi$ $\approx$ 1 for $\pi$ plasmon. The satisfactory agreement between EELS and the proposed molecular orbital model points to an obvious selection of two $\sigma$ bonds per atom that link Pb(Sn)-Se and one $\pi$ bond per Pb(Sn) to be shared by the four nearest neighbor Pb(Sn) pairs in square symmetry for all $\{$001$\}$ monoatomic layers in PSS crystal of cubic symmetry, as illustrated in Fig.~\ref{fig-PbSeMO}(d).

In addition to the evidence of $\pi$ bond existence from EELS, we find that the electron clouds along the $\langle$110$\rangle$ directions between the Pb(Sn) atoms in ED mapping (see Fig.~\ref{fig-PbSeED}(c)) can also be assigned to the high probability of $\pi$ bond electrons populated in between Pb(Sn) atoms. However, it is expected that the $\pi$ bond can only be formed one at a time among the four choices in space and time without destroying the original square symmetry.  Based on the homogeneous ED contours at the midpoint between Pb atoms along the $\langle$110$\rangle$ diagonal direction, the sharing must be evenly distributed at the four diagonal sites per $\{$001$\}$ layer in space and time.  While the electron is not dividable when fractional charge is implied, it is suggested that the $\pi$ bond has a conjugated nature similar to that observed in graphene.\cite{Chua2013}  In fact,  if the $\pi$ bond is simply randomly distributed among the four sites per $\{$001$\}$ layer permanently, i.e., the $\pi$ bond is formed without sharing as a conjugated system in space and time but randomly and permanently distributed in space only, no constructive interference from the diffraction of $\pi$ bond electrons is expected.

In summary, a molecular orbital model is proposed for Pb$_{1-x}$Sn$_x$Se, which is verified experimentally to show one unusual symmetry-protected $\pi$ bond sharing mechanism.  The identification of conjugated $\pi$ bond system in PSS compounds, similar to that observed in graphene as a gapless semiconductor with a Dirac cone shape surface band, suggests that the topological crystalline insulator Pb$_{1-x}$Sn$_x$Se has a great potential on device application via Dirac cone engineering, especially with the chemical approach on the conjugated $\pi$ bond system control.  %As a final remark, the surface mirror symmetry breaking observed in Pb$_{1-x}$Sn$_x$Se (x$\sim$0.34) at $\sim$4 K along the diagonal direction\cite{Okada2013} could also be interpreted due to the breakdown of the $\pi$ bond conjugated system at low temperature; i.e., when thermal fluctuation is reduced, a CDW-like surface reconstruction occurs following the spontaneous symmetry breaking in 2D.    

%The real space analysis of topological order has provided an intuitive and complementary understanding of the topological crystalline insulators.  
%The current results also provide an important clue to why most thermoelectric materials are also topological insulators.

%\section*{Acknowledgment}
FCC acknowledges the support provided by the Ministry of Science and Technology in Taiwan under project number MOST-102-2119-M-002 -004.  GJS acknowledges the support provided by MOST-Taiwan under project number 103-2811-M-002 -001.

%\bibliography{JabRef database} (if use this then no need to use the direct list below)

\newpage

%Below appendix two sections are part from "GJ-crystal of topological order-111114.tex" in the for words added to 6000 in this job.
%\begin{appendix}   

%\end{appendix}

\end{document}